\documentclass[12pt]{article}
\usepackage{epsfig} \usepackage{amssymb} \usepackage{amsfonts}
\def\agoth{\relax\ifmmode{\mathfrak A}\else{${\mathfrak A}${ }}\fi}
\def\pisq{\relax\ifmmode{\pi^2}\else{${\pi^2}${ }}\fi}
\def\agothk{\relax\ifmmode{\mathfrak A}_k\else{${\mathfrak A}_k${ }}\fi}
\def\acal{\relax\ifmmode{\cal A}\else{${\cal A}${ }}\fi}
\def\acalk{\relax\ifmmode{\cal A}_k\else{${\cal A}_k${ }}\fi}
\newcommand{\beq}{\begin{equation}} \newcommand{\eeq}{\end{equation}}
\newcommand{\be}{\begin{equation}}
\newcommand{\ee}{\end{equation}}
\newcommand{\bea}{\begin{eqnarray}}
\newcommand{\eea}{\end{eqnarray}}

\begin{document}
\begin{center}
{\large\bf Explicit expressions for Euclidean and Minkowskian QCD observables in analytic perturbation theory}
\medskip

{\bf D.S. Kourashev \footnote{
Moscow State University, Physical Department, Moscow, Vorobyevy Gory, 119899,
Russia\\E-mail: kourashev@mtu-net.ru
} B.A. Magradze \footnote{Bogoliubov Lab. of Theor. Physics, JINR, Dubna and A. Razmadze
 Tbilisi Mathematical Institute, 380093 Tbilisi, Georgia, E-mail: magr@rmi.acnet.ge.}
}

\medskip

{\footnotesize \bf Abstract}
\medskip

\parbox{110mm}{\footnotesize
Technical aspects of the Shirkov--Solovtsov's analytic perturbation theory (APT) are considered.
We construct explicitly two sets of specific functions, $\left\{\mathfrak{A}_n(s)\right\}$ and $\left\{{\cal A}_n(Q^2)\right\}$ that determine the nonpower asymptotic expansions for
Minkowskian and Euclidean QCD observables in APT.
The results, up to third order, are written in terms of the Lambert $W$--functions. As an input we used the exact two
loop and the three loop (corresponding to Pad\'e transformed beta-function)
RG solutions for common invariant coupling $\alpha_s$. In addition, the exact three-loop coupling is expanded in powers of the exact two-loop solution. The excellent accuracy is achieved with few terms of this series. We derive order by order elegant systems of equations for both sets of the functions.
Then we construct the global versions of the APT functions with quark thresholds in the $\overline{MS}$ scheme and give numerical results.}
\end{center}
\medskip
\section{Introduction}
In recent publications \cite{ss,ss0} a modified version of QCD perturbation theory,
free of unphysical singularities, was elaborated.
It was termed the renormalization group (RG) invariant analytic approach \cite{ss1}. In this works the idea to combine renormalization invariance and $Q^2$-analyticity for the running coupling was explored. New important properties of the analytic coupling established. This makes it possible to gain some information about infrared behavior of the theory starting from perturbative calculations.

In this paper we shall be concerned with analytic perturbation theory (APT) suggested in Ref.~\cite{mss}, a particular version of RG invariant analytic approach.
A thorough mathematical investigation of the nonpower APT series has been undertaken in works \cite{sh1}, where the stability of APT results (as compared to the conventional perturbative approach) with respect to the renormalization scheme and higher loop dependence for the whole low energy region was explained (see also \cite{ss2}).
APT has been successfully used for description of many important spacelike and timelike processes (for a partial list of contributors to this subject, see ref.~\cite{ss1}).

For the reader's convenience here we briefly summarize some aspects of APT. Let $D(Q^2)$ be a dimensionless quantity, depending on a single energy scale $Q$.
A suitable example for our purposes is Adler D-function related to some timelike process. Usually, it is presented in the form of power series
\footnote{We use the notation $Q^2=-q^2$, $Q^2>0$ corresponds to a spacelike momentum transfer.}
\be
D_{pt}(Q^2)=D_{0}(1+\sum_{n=1}^{\infty}d_{n}\alpha_{s}^{n}(Q^2,f)),
\ee
here $D_0$ is a process dependent constant and $f$ denotes the number of active quark flavors at the energy scale Q. In APT, the same quantity should be presented in the form of a nonpower asymptotic expansion as
\be
D_{an}(Q^2)=D_{0}(1+\sum_{n=1}^{\infty}d_n\acal_{n}(Q^2,f))
\ee
where ${\cal A}_n$ is the ``analyticized n-th power'' of the coupling in the spacelike region: ${\cal A}_{n}(Q^2,f)=\{\alpha_{s}^{n}(Q^2,f)\}_{an}$.
Let $R(s)$ be the physical quantity determined through $D(Q^2)$ in the timelike domain (for example $R_{e^{+}e^{-}}(s)$ or $R_\tau(M_\tau)$).
Then in APT it has the representation the \cite{ss1, sh2}
\be
\label{gth}
\begin{array}{ll}
R(s)=R_0(1+r(s)) & r(s)=\sum_{n=1}^{\infty}d_n{\agoth}_{n}(s,f).
\end{array}
\ee
The functions $\agoth_n$ are defined through the transformation 
\be
\label{R}
\agoth_{n}(s,f)=-\frac{1}{2\pi\imath}\int_{s-\imath\epsilon}^{s+\imath\epsilon}\frac{dz}{z}{\cal A}_n(-z,f),
\ee
previously introduced in Refs.~\cite{rad,kr} (see also recent paper \cite{bakul}). The inversion of (\ref{R}) reads
\be
\label{D}
{\cal A}_{n}(Q^2,f)=Q^2\int_{0}^{\infty}\frac{ds}{(s+Q^2)^2}{\agoth}_{n}(s,f).
\ee
In recent works \cite{sh1,sh2} the universal functions ${\cal A}_n$ and $\agoth_n$ have been calculated analytically at the one loop order. In the infrared region, an oscillating behavior for these functions was observed. To construct these functions, beyond the one loop order, the iterative approximation for the running coupling was used \cite{sh2,mss2,gi} or RG equation was solved numerically in the complex domain \cite{msy}.

Recently exact two-loop order solution to the RG equation has become available
\cite{my,ggk}. The solution has been written in terms of the Lambert-W function.
In addition, the third order RG equation (with Pade transformed $\beta$-function) has also been solved in terms of the same function \cite{ggk}.
These explicit solutions proved to be very useful for determination of the analyticity structure of the coupling in the complex $Q^2$-plane. Using the exact two
loop solution the analytically improved coupling ${\cal A}_{1}^{(2)}(Q^2,f)$ was reconstructed \cite{my,my1}.

Afterwards, in paper \cite{kur}, a higher order solution to the RG equation
(in arbitrary $\overline{MS}$-like renormalization scheme) was expanded in powers of the scheme independent explicit two-loop order solution. In this way, a new method for reducing the scheme ambiguity for the QCD observables has been
proposed. A similar expansion for the
observable (depending on a single energy scale Q) motivated differently, has been suggested in \cite{mx}.

The aim of this paper is to give practical formulas for calculation of the APT
sets $\left\{{\cal A}_n(Q^2,f)\right\}$ and \{$\agoth_{n}(s,f)$\}.
As an input we use the above mentioned explicit solutions to the RG equation.
In Sec. 2 we derive order by order general equations for both sets of the APT functions. In Sec. 3 we present exact two-loop results. The Minkowskian functions
are constructed, in the closed form, in terms of the Lambert-W function.
The corresponding spacelike functions, $\acal_{n}(Q^2,f)$,
are reconstructed through the spectral representation, or equivalently
using inverse transformation (\ref{D}).
In Sec. 4 the APT functions are calculated to the three loop level.
The Pade approximation improved three-loop coupling is used.
General method for constructing the APT sets, beyond two-loop level,
is discussed in Sec. 5.
In Sec. 6 we construct global universal functions, $\acal_n(Q^2)$ and $\agoth_n(s)$, (both introduced in \cite{sh2}).
For crossing the quark flavor thresholds, we use the continuous matching conditions for the $\overline{MS}$ scheme coupling
$\alpha_{s}(Q^2,f)$.
Sec. 7 contains the numerical results.
Some technical details are given in the Appendix.
The conclusions are given in Sec. 8.

\section{General results}
Let us settle conventions and notations. The running coupling of
QCD satisfies the RG equation
\be
\label{eff}
Q^2 \frac{\partial{\alpha_{s}(Q^2,f)}}{\partial{Q^2}}=
{\beta}^{f}(\alpha_{s}(Q^2,f))=-\sum_{n=0}^{\infty}
\beta_{n}^{f}\alpha_{s}^{n+2}(Q^2,f),
\ee
with the normalization condition $\alpha_{s}(\mu^{2},f)=g^2/(4\pi)$, where $\mu$ is the renormalization point.
In the class of schemes where the beta-function is mass
independent $\beta_0^{f}$ and $\beta_1^{f}$ are universal
\be
\label{coef}
\beta_{0}^{f}=\frac{1}{4\pi}\left(11-\frac{2}{3}f\right),\qquad
\beta_{1}^{f}=\frac{1}{(4\pi)^2}\left(102-\frac{38}{3}f\right),
\ee
the result for $\beta_{2}^{f}$ in the modified $MS$ $(\overline {MS})$ scheme is \cite{tvz}
\be
\beta_{2}^{f}=\frac{1}{(4\pi)^3}\left(\frac{2857}{2}-\frac{5033f}{18}+\frac{325f^2}{54}\right).
\ee
The Euclidean functions are defined through the spectral representation
\be
\label{ann}
A_n(Q^2,f)=
\frac{1}{\pi}\int_{0}^{\infty}\frac{\rho_{n}(\sigma,f)}{\sigma+Q^2}
d\sigma=
\frac{1}{\pi}\int_{-\infty}^{\infty}\frac{e^{t}}{(e^{t}+Q^2/{\Lambda}^2)}{\tilde
\rho_{n}(t,f)}dt,
\ee
where the spectral function $\rho_{n}(\sigma,f)=\Im\{\alpha_{s}(-\sigma-\imath 0)\}^n$, $\Lambda$ is the QCD scale parameter, and
$\rho_{n}(\sigma,f)\equiv \tilde
\rho_{n}(t,f)$ with $t=\ln(\sigma/\Lambda^2)$.
The timelike set of functions $\{\agoth_{n}(s,f)\}$ is defined by elegant formula
\be
\label{tr}
\agoth_{n}(s,f)
=\frac{1}{\pi}\int_{s}^{\infty}\frac{d\sigma}{\sigma}\rho_n(\sigma,f),
\ee
obtained in work \cite{ms1}.

In numerical calculations singular integral (\ref{ann}) should be regulated.
With the help of (\ref{tr}), integral (\ref{ann}) can be represented as
\be
\label{rgl}
\acal_{n}(Q^2,f)=\acal_{n}(Q^2,f,T)+\agoth_{n}(\Lambda^{2}e^{T},f)+O(e^{-T}),
\ee
where $\acal_{n}(Q^2,f,T)$ denotes the integral
(\ref{ann}) taken over the finite interval $-T\leq t\leq T$.
For sufficiently large values of $T$ (when $\exp(-T)(Q^2/\Lambda^2)\ll 1$),
the contributions of order $e^{-T}$ can be neglected. So that the second term, $\agoth_{n}(\Lambda^{2}e^{T},f)$, compensates the main truncation effects in the integral.
Formula (\ref{rgl}) enables us to achieve a good numerical precision even for moderate values of the cutoff $T$.

In what follows, we shall derive the equations for the functions $\acal_n(Q^2,f)$, $\agoth_n(s,f)$ and $\rho_n(\sigma,f)$.
In the k-loop order, the results have the form
\be
\label{rec4}
\frac{\partial \acal_{n}^{(k)}(Q^2,f)}{\partial \ln Q^2}=
-n\sum_{N=0}^{k-1}\beta_{N}^{f}\acal_{n+N+1}^{(k)}(Q^2,f),\qquad n=1.2\ldots,
\ee
\be
\label{rec5}
\frac{\partial \agoth_{n}^{(k)}(s,f)}{\partial \ln s}=
-n\sum_{N=0}^{k-1}\beta_{N}^{f}\agoth_{n+N+1}^{(k)}(s,f),\qquad n=1,2\ldots.
\ee
\be
\label{rc1}
\frac{\partial\rho_{n}^{(k)}(\sigma,f)}{\partial\ln \sigma}=-n\sum_{N=0}^{k-1}\beta_{N}^{f}\rho_{n+N+1}^{(k)}(\sigma,f), \qquad n=1,2\ldots.
\ee
The similar one-loop order equations were obtained in paper \cite{sh2}.
We start from the RG equation, in the k-loop level, presented as
\be
\label{a1}
\frac{\partial \alpha_s^{n}(Q^2,f)}{\partial \ln Q^2}=-n\sum_{N=0}^{k-1}\beta_{N}^{f}\alpha_{s}^{n+N+1}(Q^2,f)
\ee
where, $n\geq 1$, is some integer.
The APT variant of Eq.~(\ref{a1}) reads
\be
\label{a2}
\left\{\frac{\partial \alpha_s^{n}(Q^2,f)}{\partial \ln Q^2}\right\}_{an}=
-n\sum_{N=0}^{k-1}\beta_{N}^{f}{\cal A}_{n+N+1}(Q^2,f),
\ee
on the other hand, we can write
\be
\label{a3}
\left\{\frac{\partial \alpha_s^{n}(Q^2,f)}{\partial \ln Q^2}\right\}_{an}=\frac{1}{\pi}\int_{0}^{\infty}\frac{d\sigma}{\sigma+Q^2}
\Im\frac{\partial \alpha^{n}_{s}(-\sigma-\imath 0,f)}{\partial \ln(-\sigma-\imath 0)}.
\ee
From the identity, $\ln(-\sigma-\imath 0)=\ln \sigma-\imath\pi$, it follows that
\be
\label{a4}
\Im\frac{\partial \alpha^{n}_{s}(-\sigma-\imath 0,f)}{\partial \ln(-\sigma-\imath 0)}=
\frac{\partial \Im \alpha^{n}_{s}(-\sigma-\imath 0,f)}{\partial\ln\sigma}=
\frac{\partial \rho_{n}(\sigma,f)}{\partial \ln \sigma},
\ee
so that
\be
\label{byp}
\left\{\frac{\partial \alpha_s^{n}(Q^2,f)}{\partial \ln Q^2}\right\}_{an}=
\frac{1}{\pi}\int_{0}^{\infty}\frac{d\sigma}{\sigma+Q^2}\frac{\partial \rho_{n}(\sigma,f)}{\partial \ln \sigma},
\ee
integrating by parts (\ref{byp}) and using the asymptotic vanishing of
$\rho_{n}(\sigma,f)$ (the asymptotic freedom)
\be
\label{af}
\frac{\sigma}{\sigma+Q^2}\rho_n(\sigma,f)|_{0}^{\infty}=0,
\ee
 we get
\be
\left\{\frac{\partial \alpha_s^{n}(Q^2,f)}{\partial \ln Q^2}\right\}_{an}=
\frac{1}{\pi}\frac{\partial}{\partial\ln Q^2} \int_{0}^{\infty}\frac{d\sigma}{\sigma+Q^2}\rho_{n}(\sigma,f)=
\frac{\partial {\cal A}_{n}(Q^2,f)}{\partial\ln Q^2},
\ee
from this and (\ref{a2}) follows system of equations (\ref{rec4}).
From (\ref{tr}), we see that system (\ref{rc1}) is a consequence of system (\ref{rec5}).
Let us multiply Eq.~(\ref{rc1}) by the factor $(\sigma+Q^2)^{-1}$ and take the integral over the region $0<\sigma<\infty$. Integrating by parts and taking into account condition (\ref{af}), we recover the system (\ref{rec4}).
Note that for $n=1$ Eqs.~(\ref{rec4}) and (\ref{rec5})
are analogous to the basic Eq.~(\ref{eff}) with $\alpha_s^{n}$ replaced by
$\agoth_n$ and $\acal_n$ respectively.
We remark, that Eqs.~(\ref{rec4})-(\ref{rc1})
are derived on general grounds using the RG equation
(\ref{eff}) and the spectral representation together with the asymptotic freedom
condition (\ref{af}).

\section{Exact two-loop results}
For convenience, in what follows we shall omit index f in
the coefficients $\beta_n^f$.
The exact two-loop solution to Eq.~(\ref{eff}) is given by \cite{my,ggk}
\be
\label{w2}
\alpha_{s}^{(2)}(Q^2,f)=-\frac{\beta_0}{\beta_1}\frac{1}{1+
W_{-1}(\zeta)},\qquad
\zeta=-\frac{1}{eb_{1}}\left(\frac{Q^2}{\Lambda^2}\right)^{-\frac{1}{b_{1}}};
\ee
here $b_{1}=\beta_1/\beta_0^2$, $\Lambda\equiv\Lambda_{\overline {MS}}$ and $W(\zeta)$ is the Lambert W
function \cite{lamb}, the multivalued inverse of
$$\zeta=W(\zeta)\exp W(\zeta).$$
The branches of W are denoted $W_{k}(\zeta), k=0,\pm 1,\ldots .$

By performing analytical continuation of function (\ref{w2}) and its powers in the complex $Q^2$-plane \footnote{For details of analytical continuation we recommend papers \cite{my}, \cite{ggk} and \cite{my1}.}, we determine the corresponding spectral functions $\rho_n^{(2)}(\sigma,f)\equiv \tilde
\rho_{n}^{(2)}(t,f)$, $n=1,2\ldots$.
For $0\leq f \leq 6$ \footnote{For $f>6$ formula (\ref{ro}) should be changed, see \cite{ggk,my1}.}, the result is
\be \label{ro}
\tilde\rho_n^{(2)}(t,f)=\left(\frac{\beta_0}{\beta_{1}}\right)^{n}\Im\left(-\frac{1}{1+W_{1}(z(t))}\right)^n, \ee with
\be
z(t)=\frac{1}{b_{1}e}\exp(-t/b_{1}+\imath(1/b_{1}-1)\pi).
\ee
After insertion of (\ref{ro}) into (\ref{rgl}) and (\ref{tr}) we construct explicitly the two-loop functions $\acal_n^{(2)}(Q^2,f)$ and $\agoth_n^{(2)}(s,f)$.
The integrals for the Minkowskian functions can be performed \footnote{The one-loop expressions for the Minkowskian functions were derived in Refs. \cite{sh2,o}.}.
We obtain
\be
\label{f1}
\agoth_{1}^{(2)}(s,f)=-\frac{\beta_0}{\beta_1}
-\frac{1}{\pi\beta_{1}}\Im\left(\frac{1}{\alpha_{s}^{(2)}(-s)}\right)
\ee
\be
\label{f2}
\agoth_{2}^{(2)}(s,f)=\frac{1}{\pi\beta_1}\Im\ln\left(1+\frac{\beta_{1}}{\beta_0}\alpha_{s}^{(2)}(-s)\right),
\ee
\begin{eqnarray}
\label{21} {\mathfrak A}_3^{(2)}(s,f)=- \frac{\beta_0}{\beta_1}
\frac 1{\pi \beta_1} \Im \left\{ \ln \left(1+
\frac{\beta_1}{\beta_0} \alpha^{(2)}(-s)\right) -
\frac{\beta_1}{\beta_0} \alpha^{(2)}(-s)\right\}.
\end{eqnarray}
\begin{eqnarray}
\label{22} {\mathfrak A}_4^{(2)}(s,f)=\left(-
\frac{\beta_0}{\beta_1}\right)^2 \frac 1{\pi \beta_1} \Im \left\{
\ln \left(1+ \frac{\beta_1}{\beta_0} \alpha^{(2)}(-s)\right) -
\frac{\beta_1}{\beta_0} \alpha^{(2)}(-s)+
\frac{\beta_1^2}{2\beta_0^2} \alpha^{(2)2}(-s)\right\}, \qquad
etc.
\end{eqnarray}
where, for convenience, we introduced the notation
\be
\alpha^{(2)}(-s)=\alpha_{s}^{(2)}(-s-\imath 0,f)=-\frac{\beta_0}{\beta_1}\frac{1}{1+W_{1}(z_s)},
\ee
with $z_s=\frac{1}{b_{1}}(s/\Lambda^2)^{-1/b_{1}}e^{\imath\pi(1/b_1-1)-1}$.
Note that, the functions ${\mathfrak A}_n(s,f)$ are determined through $(n-2)$-th order residual terms of the Taylor expansion of
$\ln (1+\frac{\beta_1}{\beta_0} \alpha^{(2)}(-s))$
in powers of $\frac{\beta_1}{\beta_0}\alpha^{(2)}(-s)$.

Using the asymptotic properties of the W-function
\cite{lamb}, in the limit $s\rightarrow 0$, we recover the result of work \cite{ms1}
\be
\label{asy}
\begin{array}{lllll}
\agoth_{1}^{(2)}(s,f)\rightarrow \frac{1}{\beta_0},& and& \agoth_{n}^{(2)}(s,f)\rightarrow 0 & for
&n>1.\\
\end{array}
\ee
Note that,
 the ``analyticized'' nonpower perturbative expansions for the timelike observables,
may be rewritten as power series in traditional coupling
$\alpha_{s}(s)$, but with modified  by $\pi^2$-factors
coefficients. Let us consider  the two loop case.
The function (\ref{f1}) has the following formal power expansion
\be
\label{pi}
\agoth_{1}^{(2)}(s,f)=\alpha_{s}^{(2)}(s,f)-\frac{1}{3}\pi^2\beta_{0}^{2}
 \alpha_{s}^{(2)3}(s,f)-
\frac{5}{6}\pi^2\beta_{0}\beta_{1}\alpha_{s}^{(2)4}(s,f)
+\pi^2(-\frac{1}{2}\beta_1^{2}+\frac{1}{5}\pi^2 \beta_{0}^{2})\alpha_{s}^{(2)5}(s,f)+\ldots,
\ee
where $\alpha_{s}^{(2)}(s,f)$ is the exact two loop coupling (\ref{w2}). Higher order functions $\agoth_{n}$ ($n=1,2\ldots$) have similar power series expansions. Substituting these expansions into (\ref{gth}) we
find the expansion for $R(s)$ in powers of $\alpha_s$ with modified by $\pi^2$-factors coefficients.
Previously, in the timelike region, similar expansions in powers of the iterative solution to RG equation, were obtained in \cite{rad,kr}. Application of the series
can be found in
papers \cite{bjo,piv,kat}. The
  ``$\pi^2$-effects'' for various timelike quantities
have been estimated in  paper \cite{kat}, in particular, it was found that the $\pi^2$-factors give dominating contributions to the coefficients of $R_{e^{+}e^{-}}(s)=\sigma_{tot}(e^{+}e^{-}\rightarrow$ hadrons)$/\sigma(e^{+}e^{-}\rightarrow \mu^{+}\mu^{-})$. On the other hand,
in recent papers \cite{sh3,sh4} various timelike events were analyzed in the f=5 region. Higher-order ``$\pi^2$-effects'' have been taken into account properly.
It was found that the extracted values for $\alpha_{s}(M_{z}^2)$ are
influenced  significantly by these effects.

\section{The three-loop case with Pad\'e approximation}

Pad\'e improved perturbative series attracted much interest recently. They, as opposed to truncated perturbative series, exhibit reduced renormalization scale and scheme dependence \cite{gard}.
Pad\'e related resummation method which achieves elimination of the unphysical dependence from observables was proposed in recent works \cite{cvet}.

The solution to RG Eq.~(\ref{eff}), at the three-loop order,
with Pad\'e transformed beta-function
\begin{eqnarray}
\beta_{Pad\acute e}=-\beta_0
\alpha_s^2\left(1+\frac{\beta_1 \alpha_s}
{\beta_0-\frac{\beta_0 \beta_2}{\beta_1}\alpha_s}\right), \nonumber
\end{eqnarray}
has the form \cite{ggk}
\be
\label{w3}
\alpha_{Pad\acute e}^{(3)}(Q^2,f)=-\frac{\beta_0}{\beta_1}\frac{1}{1-\beta_0\beta_2/{\beta_1}^2+ W_{-1}(\xi)}:
\hspace{5mm}
\xi=-\frac{1}{eb_{1}}\exp\left(\frac{\beta_0\beta_2}{{\beta_1}^2}\right)\left(\frac{Q^2}{\Lambda^2}\right)^{-\frac{1}{b_{1}}}.
\ee
Starting from (\ref{w3}), one can construct ${\acal}^{(3)}_{Pad\acute e,n}(Q^2,f)$,
just like in the two-loop case.
For $0\leq f\leq 6$, the spectral function, $\rho_{Pad\acute e,n}(\sigma,f)\equiv {\tilde
\rho_{Pad\acute e,n}(t,f)}=\Im\{\alpha_{Pad\acute e}^{(3)n}(-\sigma-\imath 0)\}$, is defined as
\be \label{ro3}
\tilde\rho_{Pad\acute e,n}^{(3)}(t,f)=\left(\frac{\beta_0}{\beta_{1}}\right)^{n}\Im\left(-\frac{1}{1-\beta_2\beta_0/\beta_1^2+
W_{1}(Z(t))}\right)^n, \ee with
\be
Z(t)=\frac{1}{b_{1}e}\exp(\beta_0\beta_2/\beta_1^2-t/b_{1}+\imath(1/b_{1}-1)\pi).
\ee
Integral (\ref{tr}), with the spectral function (\ref{ro3}), can be done. Thus we obtain
\be
\label{frs}
\label{pd1}
\agoth_{Pad\acute e,1}^{(3)}(s,f)=-\frac{1}{\pi\beta_0}\left(\frac{1}{\eta}\Im\ln(W_{1}(Z_s))+
(1-\frac{1}{\eta})\Im\ln(\eta+W_1(Z_s))-\pi\right),
\ee

\be
\label{pd2}
\agoth_{Pad\acute e,2}^{(3)}(s,f)=\frac{1}{\pi\beta_1}\left(\frac{1}{{\eta}^2}\Im\ln\left(\frac{W_{1}(Z_{s})}{\eta+W_{1}(Z_s)}\right)
-(1-\frac{1}{\eta})\Im\left(\frac{1}{\eta+W_1(Z_s)}\right)\right),
\ee
where  $\eta=1-\beta_0\beta_2/\beta_1^2$ and
$$Z_s=\frac{1}{b}(s/\Lambda^2)^{-1/b_{1}}\exp(-\eta+\imath(1/b_{1}-1)\pi).$$
For $n>2$, the following recurrence formula is valid
\be
\label{pd3}
\agoth_{Pad\acute e,n}^{(3)}(s,f)=-\frac{\beta_0}{\eta \beta_1}\agoth_{Pad\acute e,n-1}^{(3)}(s,f)+\frac{p_n}{\eta (n-2)}\Im(\eta+W_{1}(Z_s))^{2-n}+\frac{p_{n}}{n-1}\left(\frac{1}{\eta}-1\right)\Im(\eta+W_{1}(Z_s))^{1-n},
\ee
where $p_n$ is given in (\ref{zs}). Putting together (\ref{pd2}) and (\ref{pd3}) we get
\bea
\label{pd4}
&&\agoth_{Pad\acute e,n}^{(3)}(s,f)=\frac{p_{n}}{{\eta}^{n-2}}\Im\left\{\frac{1}{\eta^{2}}\left(\ln\left(1-\frac{\eta}{\eta+W_{1}(Z_s)}\right)+\sum_{N=1}^{n-2}\left(\frac{\eta}{\eta+W_{1}(Z_s)}\right)^{N}\frac{1}{N}\right)+\right.\nonumber\\
&&\left.\frac{\eta^{n-3}(1-\eta)}{(n-1)(\eta+W_{1}(Z_s))^{n-1}}\right\}\qquad n>2.
\eea

\section{Multi-loop case}
As follows from the paper \cite{kur} one can obtain the expressions for multi-loop functions in terms of the two loop functions. In particular, the coupling of higher orders can be represented as
\be
\label{ser}
\alpha_s^{(k)}=\sum_{n=1}^{\infty}c_{n}^{(k)}\alpha_s^{(2)n},
\ee
with $c_{1}^{(k)}=1$.
The analyticized (Euclidean and Minkowskian) versions of (\ref{ser}) read
\be
\label{sum2}
{\acal}_{1}^{(k)}=\sum_{n=1}^{\infty}c_n^{(k)}{\acal}_{n}^{(2)},
\ee
\be
\label{sum3}
\agoth_{1}^{(k)}=\sum_{n=1}^{\infty}c_n^{(k)}\agoth_{n}^{(2)}.
\ee
So the two-loop functions can be considered as the minimal basis of any orders perturbative expansions \cite{kur}. Any observable $O$ (except the quantity possessing an anomalous dimension) can be represented as the series
\be
O^{(k)}=\sum_{n=1}^{\infty}O_{n}^{(k)}{\acal}_{n}^{(2)}.
\ee
We would like also to emphasize that the one-loop coupling function could
not be used for a similar expansion: the case is that multi-loop coupling functions have more complicated singularities \footnote{For example, the two-loop solution contains double logarithms
$\ln{\ln x}$ that evidently can't be expanded in powers of $1/\ln x $.}. However the exact two-loop coupling (expressed through the Lambert W-function) can be used to reflect correctly the higher orders contributions.
Note that some similar results was obtained in work \cite{mx}.

By using the limiting conditions (\ref{asy}) in the representation (\ref{sum3}), we derive immediately universal infrared limit \cite{ss,ss0} for the Minkowskian coupling to any finite order
\be
\label{lim}
\agoth_{1}^{(k)}(s,f)\rightarrow \frac{1}{\beta_0},\qquad s\rightarrow 0.
\ee
Formula (\ref{lim}) has been obtained in work \cite{ms1} in a different way.
Evidently, the same limit has the Euclidean coupling $\acal_{1}^{(k)}(Q^2,f)$ at $Q^2\rightarrow 0$ (the remarkable result of works \cite{ss,ss0}).
So we can express any observable in any loop order
in terms of the Lambert-W function. The expressions obtained will have correct
analytic properties and finite infrared limit.

\section{The quark thresholds}
In MS-like renormalization schemes important item is how to introduce
the matching conditions at the heavy
quarks thresholds for the strong coupling constant . For pedagogical introduction to this subject we recommend
paper \cite{rd} (see also Ref. \cite{btk}). In the context of APT,
this problem was studied in works \cite{sh2,sh3,sh4}.
The matching conditions and analyticity have been combined there via the special model spectral function
\be
\label{mdl}
\rho_n(\sigma)=\rho_n^{f=3}(\sigma,\Lambda_3)+\sum_{f\ge
4}\Theta(\sigma-M_f^2)(\rho_n^{f}(\sigma, \Lambda_f)-\rho_n^{f-1}(\sigma, \Lambda_{f-1})),
\ee
where the mass $M_f$ corresponds to the quark with flavor $f$
and the values for $\Lambda_f$ are determined through the continuity conditions
\be
\label{thr}
\alpha_s(M_f^2,f)=\alpha_s(M_f^2,f-1) \qquad f=4,5,6.
\ee
To be more exact, in the MS-like renormalization schemes (beyond the one loop level) conditions (\ref{thr})
should be modified \cite{rd}. With the modified conditions the final results are not sensitive to the exact scale one uses to connect the couplings. Nevertheless, in this work we assume simple formula
(\ref{thr}) up to third order: the case is that in APT this simplification does not introduce numerically significant errors.
Note that with the explicit solutions (\ref{w2}) and (\ref{w3}), the equation (\ref{thr}) for $\Lambda_f$ can be solved analytically.
For instance, inserting the three-loop solution (\ref{w3}) in (\ref{thr}) we solve
\be
\label{mth}
\Lambda_{f}=M_{f}\{-b_{f}F(z_{f-1})\exp(\eta_{f}+F(z_{f-1}))\}^{b_{f}/2}
\ee
where $b_{f}=\beta_1^f/(\beta_0^f)^2$, $\eta_{f}=1-\beta_0^f\beta_2^f/(\beta_1^f)^2$,
\be
z_{f-1}=-\frac{\exp(-\eta_{f-1})}{b_{f-1}}\left(\frac{\Lambda_{f-1}}{M_{f}}\right)^{2/b_{f-1}},
\ee
\be
F(z_{f-1})=(\eta_{f-1}+W_{-1}(z_{f-1}))\frac{\beta_1^{f-1}}{\beta_{0}^{f-1}}
\frac{\beta_0^{f}}{\beta_{1}^{f}}-\eta_{f}. \ee
Using (\ref{mdl}) with (\ref{mth}) we construct global quantities $\acal_{n}(Q^2)$ and $\agoth_{n}(s)$ relevant in the whole ranges of $Q^2$ and $s$ respectively.

\section{Numerical results}

The numerical calculations are performed by using the computer algebra system Maple V release 5; Maple has an arbitrary precision implementation of all branches of the Lambert-W function.

In Table 1, various 3-loop approximants are compared with exact three-loop coupling,
$\alpha_{num}^{(3)}$, the numerical solution of the transcendental equation
\be
\label{num}
\ln \frac{Q^2}{\Lambda^{2}}=\frac{1}{\beta_0\alpha_s}-\frac{b_{1}}{2}\ln \left(\frac{1}{\beta_{0}^{2}\alpha_s^2}+\frac{b_1}{\beta_{0}\alpha_{s} }+b_2\right)+\frac{2b_2-b_1^2}{\sqrt \Delta}\left(\arctan\frac{b_1+2b_2\beta_0\alpha_s }{\sqrt \Delta}-\arctan\frac{b_1}{\sqrt \Delta}\right),
\ee
where $b_k=\beta_{k}/\beta_{0}^{k+1}$ ($k=1,2$) and $\Delta=4b_{2}-b_1^2$.
By $\alpha_{it}^{(3)}$ we denote the commonly used three loop order iterative approximant \cite{btk}
\bea \label{it}
\alpha_{it}^{(3)}(Q^2)=\frac{1}{\beta_0
L}-\frac{\beta_1}{\beta_0^3}\frac{\ln L}{L^2}+\frac{1}{\beta_0^3
L^3}\left(\frac{\beta_1^2}{\beta_0^2}(\ln^2 L-\ln
L-1)+\frac{\beta_2}{\beta_0}\right), \eea
where $L=\ln Q^2/{\Lambda}^{2}_{\overline MS}$. Usually, the same formula (\ref{it}) is used in the
timelike region.
$\alpha_{ts}$ denotes the truncated series (\ref{ser})
\be
\alpha_{ts}^{(3)}=\sum_{i=1}^{5}c_{i}^{(3)}\alpha_s^{(2)i},
\ee
in this case $c_1^{(3)}=1$, $c_2^{(3)}=0$, $c_3^{(3)}=\frac{\beta_2}{\beta_0}$,
$c_4^{(3)}=0$ and $c_5^{(3)}=\frac{5}{3}\beta_2^2/\beta_0^2$.
From the Table 1, we see that, $\alpha_{ts}^{(3)}$ allows us to achieve more good agreement with the exact answer than the Pad\'e and iterative approximants. $\alpha_{ts}^{(3)}$
is accurate, for $Q\geq 900 $ MeV (in these region it practically coincides with the exact answer) while the Pad\'e approximant worsens below $Q=1600 MeV$ and iterative approximant allows us to achieve $1\%$ accuracy only for $Q\geq 2300$ $MeV$.
We have taken $\Lambda_3=400$ $MeV$, and the values of $\Lambda_f$ $(f=4,5,6)$ are determined by imposing conditions (\ref{thr}) on the exact numerical solution (\ref{num}). For the quarks masses throughout  this paper we assume the values
$M_1=M_2=M_3=0$, $M_4=1.3$ $GeV$, $M_5=4.3$ $GeV$ and $M_6=170$ $GeV$.

In Table 2 the results for the Minkowskian expansion functions are given in the three quark region (0.4 GeV$<{\sqrt s}<$2.6 GeV). The functions $\agoth^{(3)}_{Pade,n}(s,f)$, $\agoth^{(2)}_{n}(s,f)$ and $\agoth^{(3)}_{ts,n}(s,f)$ are compared for $n=1,2$ and $f=3$.
We take the common value, $\Lambda_{f=3}=400$ MeV. The differences between $\agoth_{ts,1}^{(3)}$
and $\agoth_{1}^{(2)}$ are less than 2.19\%, while the differences between
$\agoth_{ts,1}^{(3)}$ and $\agoth_{Pad{\acute e},1}^{(3)}$ are smaller than 0.12\%. For the second functions $(n=2)$, these differences are of the same order: $|\delta\agoth_2^{(2)}(\%)|<2.6$ and $|\delta\agoth_{Pade,2}^{(3)}(\%)|<0.3$.

Table 3 summarizes the results for Euclidean expansion functions, $\acal_n^{(k)}(Q^2,f)$, in the two loop and three-loop orders, for $n=1,2$ and $f=3$. The three quark region is chosen (0.4 GeV $<Q<$ 2 GeV) and we take $\Lambda_{f=3}=400$ MeV. The numerical calculations have been performed using formula (\ref{rgl})
(high precision was achieved already for $T=1000$).
In the three loop case, two approximants, $\acal_{ts,n}^{(3)}$ and $\acal_{Pad\acute e,n}^{(3)}$, are compared. There is good agreement between these approximants ( 0.07\% end better for n=1, and 0.14\% and better for $n=2$).
The differences between $\acal_{ts,n}^{(3)}$ and $\acal_{n}^{(2)}$ $(n=1,2)$
are larger, up to 1.72\% for $n=1$ and up to 1.4\% for $n=2$.

The asymmetry \cite{ms4}, $\delta_{as}(\%)=(\acal_{tr,1}^{(3)}(Q^2,f)-\agoth_{tr,1}^{(3)}(Q^2,f))/\acal_{tr,1}^{(3)}(Q^2,f)*100$, was found to be appreciable (compare Tables 2 and 3): it increases from 2\% at $Q=\sqrt s=0.4$ $GeV$ to 7.5\% at $Q=\sqrt s=2.0$ $GeV$.

On the other hand, for global expansion functions $\acal_{n}(Q^2)$
and $\agoth_n(s)$ (that are constructed according the model (\ref{mdl})) all the above mentioned differences are slightly enhanced.
In Table 5, the results obtained with Minkowskian global approximant, $\agoth_{ts,1}^{(3)}(s)$, are given. For comparison we include the differences,
$d^{(2)}(\%)=(\agoth_{ts,1}^{(3)}(s)-\agoth_{1}^{(2)}(s))/\agoth_{ts,1}^{(3)}(s)*100$ and
$d^{(3)}(\%)=(\agoth_{ts,1}^{(3)}(s)-\agoth_{Pad\acute e,1}^{(3)}(s))/\agoth_{ts,1}^{(3)}(s)*100$.
As an input we take $\Lambda_{f=3}=400$ MeV. The values of $\Lambda_f$ ($f=4,5,6$) for various approximants (determined through (\ref{thr})) are given in Table 4.
The maximal differences are observed at $\sqrt s\approx 1.3$ $GeV$: $d^{(2)}(\%)=3.21$ and $d^{(3)}(\%)=-0.63$ respectively.
The analogical results for the global Euclidean coupling $\acal_{1}(Q^2)$ are given in Table 6.
In this case $D^{(2)}(\%)<2.64$ and $|D^{(3)}(\%)|<0.5$ for 0.4 GeV $<Q<$ 100 GeV.

\section{Conclusions}
In this paper we have constructed analytically two sets of the specific functions, $\{\acal_n\}$ and $\{\agoth_n\}$, that determine the APT expansions for the spacelike and timelike observables respectively. The results, up to third order, are written in terms of the Lambert W-function. For the both sets of the functions,
we have derived order by order the infinite sets of differential equations.
These equations follow from the RG equation, spectral representation and asymptotic freedom. With the help of the equations, at the two-loop level, compact expressions for the Minkowskian functions $\agoth_n(s,f)$ (see Eqs.(\ref{f1})-(\ref{22})) are obtained.
For the Euclidean functions, $\acal_n(Q^2,f)$, the regulated expressions (\ref{rgl}) guaranteeing high precision in numerical calculations are presented.
We have shown the connection between APT series for the timelike observables and the corresponding power series with modified (by ``$\pi^2$-factors'') coefficients (see Eq.~(\ref{pi})).

For Pade-improved 3-loop case, the Minkowskian set of functions $\{\agoth_{n}(s,f)\}$ is constructed (see (\ref{pd4})).
In addition, to calculate the APT functions in the three-loop order, the general method of work \cite{kur} is used. The advantage of this method over the method of Pad\'e approximation is demonstrated (see Table 1).

The matching conditions for the running coupling function $\alpha_{s}(Q^2,f)$
at the flavor thresholds to three loops in $\overline {MS}$ scheme are presented in analytic form (see (\ref{mth})).
Global functions $\acal_{n}(Q^2)$ and $\agoth_{n}(s)$ with flavor thresholds are constructed.

We have examined numerically some of the APT functions to second and third orders (see Tables 1-6). The accuracy of various approximants to the spacelike and timelike APT functions is estimated.

Let us discuss the status of the obtained exact and approximate expressions for
the functions $\acal_n$ and $\agoth_n$. The question
may be raised as to whether they allow us to achieve advantage over the existing in the literature iterative formulas. In recent works \cite{bakul,mss2,gi}, analytical continuation to the timelike region
has been performed using the iterative solution (\ref{it}) to the RG equation.
However, it was estimated in ref.~\cite{myu},
 that with the iterative solution, in the two loop case,
the error in $\acal_1^{(2)}(Q^2,f)$ and $\agoth_1^{(2)}(s,f)$ may be as large as 4\%-5\%. Furthermore, the iterative formulas are not convenient
from the technical point of view too. Since, starting from iterative formulas, we have not found analogous of Eqs.(\ref{rec4})-(\ref{rec5}) which greatly simplify numerical analysis.
On the other hand, the precision of the experimental data (for $R_{e^{+}e^{-}}(s)$, $R_{\tau},$ ets.) is continuously increasing. Therefore the authors hope that the more accurate theoretical formulas will be helpful for practical calculations in the APT framework.
It should be remarked, that an alternative to our method was given in work \cite{msy}, where the RG equation, to third order, has been solved numerically in the complex domain.

Other possible applications of the obtained formulas are in the context of different (not APT) approaches to Landau pole problem \cite{kp, nau, nst} and more general (non-perturbative) frameworks developed in works \cite{aa, dmw, gr}.\\

{\bf Acknowledgments}\\
The authors would like to thank D.V. Shirkov for numerous valuable discussions.
We are grateful to A.L. Kataev, A.A. Pivovarov, A.V. Sidorov, I.L. Solovtsov, O.P. Solovtsova
for critical comments and stimulating discussions.

\section{Appendix}
In this appendix we calculate analytically $\agoth_{n}(s,f)$ to second order. Let us define the auxiliary functions
\be
\label{ox} R_n(s,f)=\frac{1}{\pi}\int_{\ln\bar s}^{\infty}dt{\tilde
a}^n(t,f).
 \ee
where $\bar s=s/\Lambda^2$ and
 $\Im {\tilde a}^n(t,f)={\tilde \rho}_{n}(t,f)\equiv\rho_{n}(\sigma,f)$ with $\sigma=\exp(t)$.
Then $\agoth_n(s,f)=\Im R_n(s,f).$
The expressions for $\tilde a$, can be read
from (\ref{ro}).
In the two loop case
\be
\label{2l}
{\tilde a}^{(2)}(t,f)=-\frac{\beta_0}{\beta_1}\displaystyle\frac{1}{1+W_1(z(t))}.
\ee
With (\ref{2l}) integral (\ref{ox}) can be
rewritten as a contour integral in the complex z-plane
\be
\label{oxx}
R_n^{(2)}(s,f)=p_n\int_{z_\epsilon}^{z_{s}}
\frac{dz}{z}\frac{1}{(1+W_{1}(z))^n}, \ee where, we denote
\be
\label{zs}
p_n=\frac{1}{\pi\beta_1}\left(-\frac{\beta_0}{\beta_1}\right)^{n-2},\qquad
z_s=z(\ln s),\qquad
z_{\epsilon}=\epsilon e^{\imath (\frac{1}{b_{1}}-1)},
\ee
and the limit
$\epsilon\rightarrow 0$ is assumed.
Let us change the integration variable in (\ref{oxx}), $\omega=W(z)$, we then get
\be
\label{o}
R_n^{(2)}(s,f)=p_n\int_{W_{1}(z_{\epsilon})}^{W_{1}(z_{s})}
\frac{1+\omega}{\omega(1+\omega)^n}d\omega.\ee For $n>2$,
from (\ref{o}) we obtain the equation
\be
\label{atn1}
{\agoth}_{n}^{(2)}(s,f)=-\frac{\beta_0}{\beta_1}{\agoth}_{n-1}^{(2)}(s,f)+\frac{p_n}{(n-2)}\Im(1+W_{1}(z_s))^{2-n},\qquad n=3,4\ldots,
\ee
the two-loop version of the equation (\ref{rec5}).
It is sufficient, to calculate
$\agoth_1^{(2)}$ and $\agoth_2^{(2)}$. Indeed, higher functions can be determined using recurrence formula (\ref{atn1}). Note that  $R_1^{(2)}(s,f)$ is divergent
in the limit $\epsilon\rightarrow 0$, (see (\ref{oxx}) ).  Nevertheless,  it has finite imaginary
part \footnote{For the asymptotic behaviour of the W function see paper
\cite{lamb}.}. By direct calculation we obtain formulas (\ref{f1}) and (\ref{f2}), formulas (\ref{21}) and (\ref{22}) follow from formula (\ref{f2}).

\begin{center}
\begin{table}[h]
\caption{The various approximations to the three loop coupling versus exact numerical solution ${\alpha}_{num}^{(3)}$ of the RG equation.
We include the relative errors $\delta{\alpha_{ts}^{(3)}(\%)}=(\alpha_{num}^{(3)}-\alpha_{ts}^{(3)})/\alpha_{num}^{(3)}*100$,
$\delta\alpha_{Pade}^{(3)}(\%)=(\alpha_{num}^{(3)}-\alpha_{Pade}^{(3)})/\alpha_{num}^{(3)}*100$ and
$\delta\alpha_{it}^{(3)}(\%)=(\alpha_{num}^{(3)}-\alpha_{it}^{(3)})/\alpha_{num}^{(3)}*100$.}
\vspace{0.2cm}
\begin{center}

\begin{tabular}{lllll}
\hline
$Q$ $ GeV $   &
$  {\alpha}^{(3)}_{num}(Q^2,f)$ &
$\delta\alpha_{ts}^{(3)}(\%)$&
$\delta\alpha_{Pade}^{(3)}(\%)$&
$\delta\alpha_{it}^{(3)}(\%)$\\ \hline
\multicolumn{5}{c}{$f=3$ \hspace{5mm}$\Lambda_3=400$ $MeV$}\\
\hline
 .8   &   .764911     &       1.90   &    -18.88    &       -15.5      \\
.9    &   .633231     &      .82     &   -7.42       &     -9.3        \\
1.0  &    .554140    &     .43      &    -4.40       &     -6.1        \\
1.1  &    .500275    &    .26       &    -3.04       &     -4.3        \\
1.2  &    .460747    &   .17        &    -2.28       &     -3.3        \\
1.3  &    .430253    &  .12        &    -1.81       &     -2.6         \\
1.4  &    .405868    &    .09       &    -1.49       &     -2.2        \\
1.5  &    .385829    &    .06       &    -1.26       &     -1.9        \\
1.6  &    .369009    &     .05       &    -1.09      &      -1.6       \\
1.7  &    .354645    &      .04      &    -.96       &      -1.5       \\
1.8  &    .342204    &      .03      &    -.85       &      -1.3       \\
1.9  &    .331300    &      .03    &    -.77      &       -1.2      \\
2.0  &    .321646  &      .02       &    -.70       &      -1.1       \\
2.2  &    .305269  &      .02       &    -.59       &      -1.0       \\
2.4  &    .291842  &     .01        &    -.51       &      -.9         \\
2.6 .&    .280585 .&     .01       .&    -.45       &    -.9           \\ \hline
\multicolumn{5}{c}{$f=4$ \hspace{5mm}$\Lambda_4=354.407$ $MeV$}\\
\hline
2 &  .330496  &  .009 &    -.40  &    -.39     \\
4 & .244183  &   .002&    -.15 &     -.29    \\
6 & 212689  &   .001&    -.10 &     -.26    \\
8 & 195076  &   .0004&    -.08 &     -.24    \\
10& 183390  &   .0003&     -.06&     -.23    \\      \hline
\multicolumn{5}{c}{$f=5$ \hspace{5mm}$\Lambda_5=259.602$ $MeV$}\\ \hline
10  &  .187687   &  .00003 & -.02 & -.01 \\
20  &  .160349   &  .00001 & -.01 & -.04  \\
30  &  .147853   &  .00001 & -.01 & -.04  \\
\hline
\end{tabular}
\end{center}
\end{table}
\end{center}

\begin{center}
\begin{table}[h]
\caption{Comparison of the Minkowskian functions $\agoth_{n}^{(2)}(s,f)$, $\agoth_{Pade,n}^{(3)}(s,f)$ and $\agoth_{ts,n}^{(3)}(s,f)$. We denote $\delta\agoth_{n}^{(2)}(\%)=
(\agoth_{ts,n}^{(3)}-\agoth_{n}^{(2)})/ \agoth_{ts,n}^{(3)}*100$ and
${\delta}\agoth_{Pad{\acute e},n}^{(3)}(\%)=
(\agoth_{ts,n}^{(3)}-\agoth_{Pad{\acute e},n}^{(3)})/ \agoth_{ts,n}^{(3)}*100$.}
\vspace{0.2cm}

\begin{center}
\begin{tabular}{lllllll}
\hline
${\sqrt s}$  $ GeV $   &
$  {\agoth}_{ts,1}^{(3)}(s,3)$ &
${\delta}\agoth_{1}^{(2)}(\%)$&
${\delta}\agoth_{Pad{\acute e},1}^{(3)}(\%)$& ${\agoth}_{ts,2}^{(3)}(s,3)$ & $\delta\agoth_{2}^{(2)}(\%)$& $\delta\agoth_{Pad{\acute e},2}^{(3)}(\%)$ \\ \hline
\multicolumn{7}{c}{$f=3$ \hspace{5mm}$\Lambda_3=400$ $MeV$}\\
\hline
.4 &  .501609 &  1.27 &  .11 & .130889  & -1.99 & .28 \\
.5 &  .458689 &  1.65 &  .07 &.122465   & -1.03 & .30 \\
.6 &  .425397 &  1.90 &  .03 &.114203   & -.18  & .29  \\
.7 &  .398821 &  2.05 &  -.01&.106619   & .51   &.25  \\
.8 &  .377110 &  2.14 &  -.04&.099846   & 1.04  &.21  \\
.9 &  .359032 &  2.18 &  -.06&.093863   & 1.45  &.16  \\
1.0&  .343733 &  2.19 &  -.08&.088593   & 1.75  &.12  \\
1.2&  .319204 &  2.16 &  -.10&.079832   & 2.16  &.05  \\
1.4&  .300337 &  2.11 &  -.11&.072913   & 2.39  &-.01   \\
1.6&  .285311 &  2.04 &  -.12&.067343   & 2.51  &-.04   \\
1.8&  .273011 &  1.98 &  -.12&.062776   & 2.57  &-.07   \\
2.0&  .262720 &  1.91 &  -.12&.058965   & 2.60  &-.09   \\
2.2&  .253955 &  1.85 &  -.12&.055737   & 2.60  &-.10   \\
2.4&  .246378 &  1.79 &  -.12&.052967   & 2.59  &-.11   \\
2.6&  .239747 &  1.74 &  -.11&.050561   & 2.57  &-.12   \\
\hline
\end{tabular}
\end{center}
\end{table}
\end{center}

\begin{center}
\begin{table}[h]
\caption{Comparison of the Euclidean functions, $\acal_n^{(k)}(Q^2,f)$, at the two-loop and three loop orders.}
\vspace{0.2cm}
\begin{center}

\begin{tabular}{lllllll}
\hline
$Q$  $ GeV $   &
$  {\acal}_{1}^{(2)}(Q^2,3)$ &
${\acal}_{Pad\acute e,1}^{(3)}(Q^2,3)$&
${\acal}_{ts,1}^{(3)}(Q^2,3)$&
${\acal}_{2}^{(2)}(Q^2,3)$ &
 ${\acal}_{Pad\acute e,2}^{(3)}(Q^2,3)$&
${\acal}_{ts,2}^{(3)}(Q^2,3)$ \\ \hline
.4 & .507853 & .511591  & .511933 &   .118913  &  .117136   & .117273 \\
.6 & .438444 & .443587  & .443761&    .107177  &  .106285   & .106436 \\
.8 &  .393408 & .399067  & .399119&    .097030  &  .096780   & .096911 \\
1.0& .361380&  .367164 &  .367133&    .088705 &   .088884  &  .088990 \\
1.2& .337219&  .342947 &  .342862&    .081875 &   .082339  &  .082421 \\
1.4& .318218&  .323809 &  .323686&    .076208 &   .076860  &  .076922 \\
1.6& .302807&  .308224 &  .308076&    .071443 &   .072219  &  .072264 \\
1.8& .290003&  .295233 &  .295069&    .067385 &   .068241  &  .068273 \\
2.0& .279159&  .284203 &  .284028&    .063886 &   .064794  &  .064815 \\
2.2&   .269831&  .274693 & .274511&  .060839  & .061779 &  .061790 \\
2.4&   .261702&  .266390 & .266205&  .058160  & .059117 &  .059121 \\
2.6&   .254538&  .259063 & .258876&  .055784 &  .056749&   .056747 \\
\hline
\end{tabular}
\end{center}
\end{table}
\end{center}

\begin{center}
\begin{table}[h]
\caption{The flavor dependence of the ${\overline {MS}}$ scheme scale $\Lambda_{f}$, as determined from the approximants $\alpha_s^{(2)}$, $\alpha_{num}^{(3)}$, $\alpha_{ts}^{(3)}$ and $\alpha_{Pad\acute e}^{(3)}$.}
\vspace{0.2cm}
\begin{center}

\begin{tabular}{c|cccc}
\hline
f&
3&
4&
5&
6\\ \hline
$\Lambda_f^{(2)}$ $MeV$&400&   341.444&   242.195&   102.113\\
$\Lambda_{num,f}^{(3)}$ $MeV$&400& 354.407&   259.602&   112.194\\
$\Lambda_{ts,f}^{(3)}$ $MeV$&400& 354.034&   259.288&  112.043\\
$\Lambda_{Pade,f}^{(3)}$ $MeV$&400&   358.667&  263.932&   114.306\\

\hline
\end{tabular}
\end{center}
\end{table}
\end{center}

\begin{center}
\begin{table}[h]
\caption{Comparison of various approximants for global Minkowskian coupling $\agoth_{1}(s)$. $d^{(2)}(\%)$ and $d^{(3)}(\%)$ denote relative errors
for $\agoth_{1}^{(2)}(s)$ and $\agoth_{Pad\acute e,1}^{(3)}(s)$ respectively.}
\vspace{0.2cm}
\begin{center}

\begin{tabular}{llll|l|llll}
\hline
$\sqrt s$  $ GeV $   &
$  {\agoth}_{ts,1}^{(3)}(s)$ &
$d^{(2)}(\%)$&
$d^{(3)}(\%)$ &
&
$\sqrt s$  $ GeV $ &
$  {\agoth}_{ts,1}^{(3)}(s)$ &
$d^{(2)}(\%)$&
$d^{(3)}(\%)$ \\ \hline
.4 &  .520162 &   1.98 & -.23&& 3  &.245069   &  2.61 & -.54   \\
.5 &  .477242 &   2.41 & -.30&& 4  &.224780   &  2.42 & -.51   \\
.6 &  .443950 &   2.70 & -.36&& 5  &.211545   &  2.27 & -.48\\
.7 &  .417373 &   2.90 & -.42&& 6  &.201948   &  2.17 & -.46\\
.8 &  .395663 &   3.03 & -.47&& 7  &.194467   &  2.08 & -.45\\
.9 &  .377585 &   3.11 & -.51&& 8  &.188409   &  2.01 & -.43\\
1.0&   .362286&   3.16 & -.55&& 9  &.183363   &  1.95 & -.42\\
1.1&   .349158&   3.19 & -.58&& 10 & .179068  &   1.90&   -.40\\
1.2&   .337757&   3.21 & -.60&& 20 & .155103  & 1.63 & -.36\\
1.3&   .327751&   3.21 & -.63&& 25 & .148685  & 1.55 & -.34\\
1.4&   .318641&   3.17 & -.62&& 30 & .143821  & 1.50 & -.33\\
1.6&   .303211&   3.08 & -.61&& 50 & .131743  & 1.36 & -.30\\
1.8&   .290593&   2.99 & -.60&& 70 & .124838  &  1.28&   -.28\\
2.0&   .280039&   2.91 & -.59&& 90 & .120137  &  1.23&   -.27\\
2.4&   .263277&   2.77 & -.57&&100 &  .118270 & 1.20&   -.27\\

\hline
\end{tabular}
\end{center}
\end{table}
\end{center}

\begin{center}
\begin{table}[h]
\caption{Comparison of various approximants to global Euclidean function
$\acal_{1}(Q^2)$. The relative errors $D^{(2)}(\%)$ and $D^{(3)}(\%)$ correspond to $\acal_{1}^{(2)}(Q^2)$ and $\acal_{1,Pade}^{(3)}(Q^2)$ respectively.}
\vspace{0.2cm}
\begin{center}

\begin{tabular}{llll|l|llll}
\hline
$Q$  $ GeV $   &
$  {\acal}_{ts,1}^{(3)}(Q^2)$ &
$D^{(2)}(\%)$&
$D^{(3)}(\%)$ &
&
$Q$  $ GeV $ &
$  {\acal}_{ts,1}^{(3)}(Q^2)$ &
$D^{(2)}(\%)$&
$D^{(3)}(\%)$ \\ \hline
.4  &  .530379 &  1.50 &-.25 &   &  7   & .205366 & 2.22& -.46 \\
.6  &  .462088 &  1.97 &-.32 &   &  8   & .198339 & 2.15& -.45 \\
.8  &  .417301 &  2.25 &-.37 &   &  9   & .192520 & 2.10& -.44 \\
1   &  .385158 &  2.42 &-.41 &   &  10  & .187592 & 2.05& -.43 \\
1.2 &  .360725 &  2.53 &-.44 &   &  20  & .160570 & 1.75& -.38 \\
1.4 &  .341394 &  2.59 &-.46 &   &  30  & .148141 & 1.60& -.35 \\
1.6 &  .325635 &  2.62 &-.48 &   &  40  & .140458 & 1.50& -.33 \\
1.8 &  .312489 &  2.64 &-.49 &   &  50  & .135046 & 1.44& -.32 \\
2   &  .301321 &  2.64 &-.49 &   &  60  & .130940 & 1.39& -.31 \\
3   &  .263297 &  2.58 &-.50 &   &  70  & .127669 & 1.35& -.30 \\
4   &  .240696 &  2.48 &-.50 &   &  80  & .124975 & 1.32& -.29 \\
5   &  .225362 &  2.38 &-.48 &   &  90  & .122699 & 1.29& -.29 \\
6   &  .214096 &  2.29 &-.47 &   &  100 & .120740 & 1.26& -.28 \\

\hline
\end{tabular}
\end{center}
\end{table}
\end{center}

\end{document}